\documentclass[article,twocolumn,prd,superscriptaddress]{revtex4}

\usepackage{amssymb}
\usepackage{amssymb}
\usepackage{amsmath}
\usepackage{feynmf}
\usepackage{slashed}
\usepackage{natbib}
\usepackage{graphicx}
\usepackage[pdftex,bookmarks,linktocpage,pdfpagelabels,plainpages=false,hyperfigures,linkcolor=blue,citecolor=blue]{hyperref} 
\hypersetup{colorlinks=true,allcolors=blue}

\usepackage{mathrsfs,amssymb}  
\usepackage{cancel}
\usepackage[normalem]{ulem}

\newcommand\be{\begin{equation}}
\newcommand\ee{\end{equation}}

\newcommand\bea{\begin{eqnarray}}
\newcommand\eea{\end{eqnarray}}

\begin{document}

\bibliographystyle{apsrev4-1}

\title{Dark Matter-Neutrino Interconversion at \\COHERENT, Direct Detection, and the Early Universe}

\author{Nicholas Hurtado}
\affiliation{Center for Neutrino Physics, Department of Physics, Virginia Tech University, Blacksburg, VA 24601, USA}

\author{Hana Mir}
\affiliation{Center for Neutrino Physics, Department of Physics, Virginia Tech University, Blacksburg, VA 24601, USA}

\author{Ian M. Shoemaker}
\affiliation{Center for Neutrino Physics, Department of Physics, Virginia Tech University, Blacksburg, VA 24601, USA}

\author{Eli Welch}
\affiliation{Center for Neutrino Physics, Department of Physics, Virginia Tech University, Blacksburg, VA 24601, USA}

\author{Jason Wyenberg}
\affiliation{Department of Physics, University of South Dakota, Vermillion, SD 57069, USA}

\begin{abstract}

We study a Dark Matter (DM) model in which the dominant coupling to the standard model occurs through a neutrino-DM-scalar coupling. The new singlet scalar will generically have couplings to nuclei/electrons arising from renormalizable Higgs portal interactions. As a result the DM particle $X$ can convert into a neutrino via scattering on a target nucleus $\mathcal{N}$: $ X + \mathcal{N} \rightarrow \nu + \mathcal{N}$, leading to striking signatures at direct detection experiments. Similarly, DM can be produced in neutrino scattering events at neutrino experiments: $ \nu + \mathcal{N} \rightarrow X + \mathcal{N}$, predicting spectral distortions at experiments such as COHERENT. Furthermore, the model allows for late kinetic decoupling of dark matter with implications for small-scale structure. At low masses, we find that COHERENT and late kinetic decoupling produce the strongest constraints on the model, while at high masses the leading constraints come from DM down-scattering at XENON1T and Borexino. Future improvement will come from  CE$\nu$NS data, ultra-low threshold direct detection, and rare kaon decays.

\end{abstract}

\maketitle

\section{Introduction}
The most abundant type of matter in the Universe remains unknown. While the existence of this ``Dark Matter" (DM) is supported by a number of observations, they are all gravitational in nature and do not provide information about the particle nature of DM. Given the enormous variety of experimental activity in the search for DM, there is hope that future data will clarify the particle characteristics of DM.

The observation of neutrino flavor oscillations, implying the existence of neutrino masses, is also not predicted by the Standard Model. As such, neutrinos also require new physics. It is therefore natural to consider models in which neutrinos and DM share new interactions. Models of ``neutrinophilic'' dark matter induce novel modifications of the power spectrum, and may have important implications for small-scale structure~\cite{Boehm:2003hm,Hooper:2007tu,Aarssen:2012fx,Dasgupta:2013zpn,Shoemaker:2013tda,Cherry:2014xra,Bertoni:2014mva,Binder:2016pnr,Campo:2017nwh}, induce modifications in high-energy neutrino fluxes~\cite{Davis:2015rza,Cherry:2016jol,Arguelles:2017atb,Kelly:2018tyg,Farzan:2018pnk,Pandey:2018wvh,Choi:2019ixb}, solar neutrinos~\cite{Capozzi:2017auw}, atmospheric neutrinos~\cite{Capozzi:2018bps} , and these DM-neutrino interactions may even provide a route for explaining the observed DM abundance via the thermal freeze-out mechanism~\cite{Scherrer:1985zt} (e.g.,~\cite{Boehm:2003hm,Hooper:2007tu,Cherry:2014xra,Arguelles:2019ouk}). 

In this paper we study the effects of a new interaction between DM and neutrinos mediated by a scalar $\phi$ via the Yukawa interaction, $\mathcal{L} \supset y_{X} \bar{X} \phi \nu$, where the scalar also couples to nuclei. In the presence of this new interaction, DM can convert into neutrinos upon scattering on nuclei. This interaction leads to novel recoil spectra at DM direct detection and neutrino experiments and has recently been studied by Dror, Elor, and McGehee~\cite{Dror:2019onn,Dror:2019dib}. The reverse process also exists, which allows incoming neutrinos to convert into DM when incident on a target nucleus~\cite{Brdar:2018qqj}. We study the implications of these search strategies at the COHERENT~\cite{Akimov:2017ade}, XENON1T~\cite{Aprile:2018dbl}, and Borexino experiments~\cite{Bellini:2013uui} and find that they nearly rule out the thermal relic hypothesis for the DM abundance in the model. 

The remainder of this paper is organized as follows. In the next section, we introduce the details of the model and discuss some of the baseline phenomenological constraints. In Sec.~\ref{sec:COH} we study the production of DM from neutrinos scattering off nuclei at COHERENT. In Sec.~\ref{sec:KD} we examine the constraints coming from the late kinetic decoupling between DM and neutrinos.  In Sec.~\ref{sec:DD} we discuss direct detection signatures and constraints. In Sec.~\ref{sec:disc} we discuss the complementarity of these bounds taken together and examine the dependence on the mediator mass. Finally in Sec.~\ref{sec:conc} we conclude and mention future prospects.

\begin{figure*}[t!]
\includegraphics[angle=0,width=.99\textwidth]{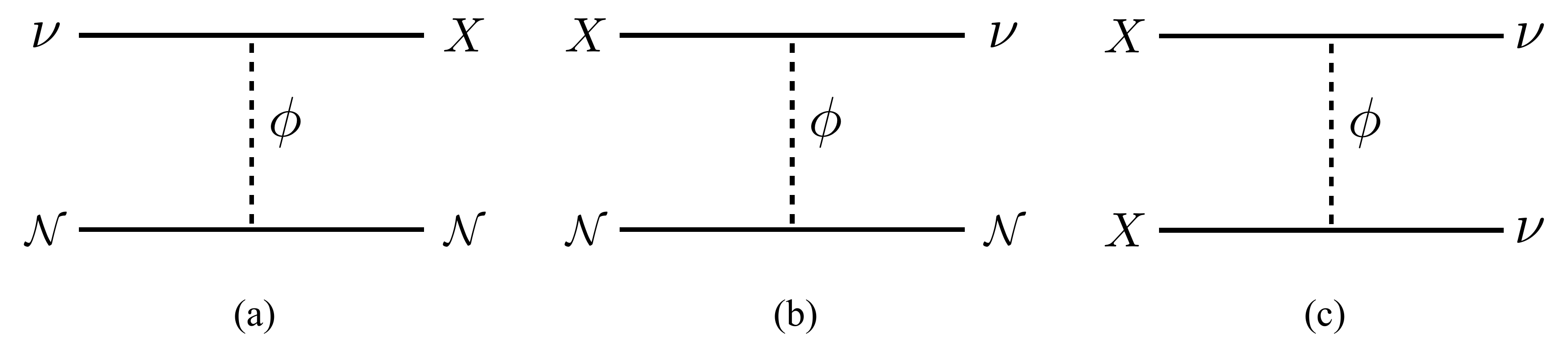}
\caption{Lowest order Feynman diagrams describing interactions between nuclei $\mathcal{N}$, dark matter $X$ and neutrinos $\nu$. The left panel (a), describes neutrino-to-DM conversion upon scattering on a nucleus $\mathcal{N}$, the middle panel (b) depicts DM-to-neutrino conversion upon scattering on a nucleus, and the right panel (c) DM pair annihilation to a pair of neutrinos mediated by the scalar $\phi$. Process (a) leads to constraints from neutrino experiments such as COHERENT, process (b) determines direct detection constraints, and the $X+\nu \leftrightarrow X+\nu$ scattering in (c) determines the kinetic decoupling of dark matter. }
\label{fig:diagrams}
\end{figure*}

\section{Model Setup}
\label{sec:model}
We study the following simple model in which the DM particle $X$ is a fermion that interacts with neutrinos and a new scalar mediator $\phi$, via the Lagrangian
\be
\label{eq:model2}
\mathcal{L} \supset y_{X} \bar{X} \phi \nu + y_{q} \phi \bar{q}q,
\ee
where the Yukawa couplings $y_{X}$ and $y_{q}$ control the strength of the DM-neutrino-$\phi$ and quark-$\phi$ interactions respectively. In this paper we focus on the implications of this interaction for terrestrial experiments, and note that possible UV completions for the model have been discussed in Ref.~\cite{Brdar:2018qqj}.

It is well-known that the coupling of $\phi$ to SM fermions is constrained by a variety of terrestrial experiments (e.g.~Ref.~\cite{Krnjaic:2015mbs,Batell:2018fqo}). In the mass range of interest some of the strongest bounds come from $K$ and $B$ meson decays. This is due to the fact that the quark couplings in Eq.~(\ref{eq:model2}) allow for the production of the new scalar via $B^{+} \rightarrow K^{+} + \phi$ or $K^{+} \rightarrow  \pi^{+}+ \phi$, while the DM-$\nu$ coupling allows for $\phi$ to decay invisibly $\phi \rightarrow \bar{\nu} + X$. In our mass range of interest, the kaon decay measurements of $K \rightarrow \pi + \bar{\nu} +\nu$ from the E787 and E949 experiments at Brookhaven National Laboratory~\cite{Artamonov:2008qb} set the strongest limits. At higher $\phi$ masses BaBar's constraints on $B^{+} \rightarrow K^{+} +$ invisible sets the leading constraint on $y_{q}$.

We note that the NA62 experiment at CERN's Superproton Synchrotron aims to improve the bounds on $K^{+} \rightarrow \pi^{+} \bar{\nu} \nu$ by measuring the branching ratio to 10$\%$ precision~\cite{CortinaGil:2018fkc,Lurkin:2019brq}. This roughly corresponds to an order of magnitude improvement in the bound on a BSM contribution to the branching ratio over what is allowed by E787 and E949~\cite{Artamonov:2008qb}. So far NA62 has only published results using 2$\%$ of the data accumulated through 2018~\cite{CortinaGil:2018fkc}, and thus new bounds on ${\rm BR}(K^{+} \rightarrow \pi^{+} \bar{\nu} \nu)$ may be imminent.

\begin{figure*}[t]
\includegraphics[angle=0,width=.48\textwidth]{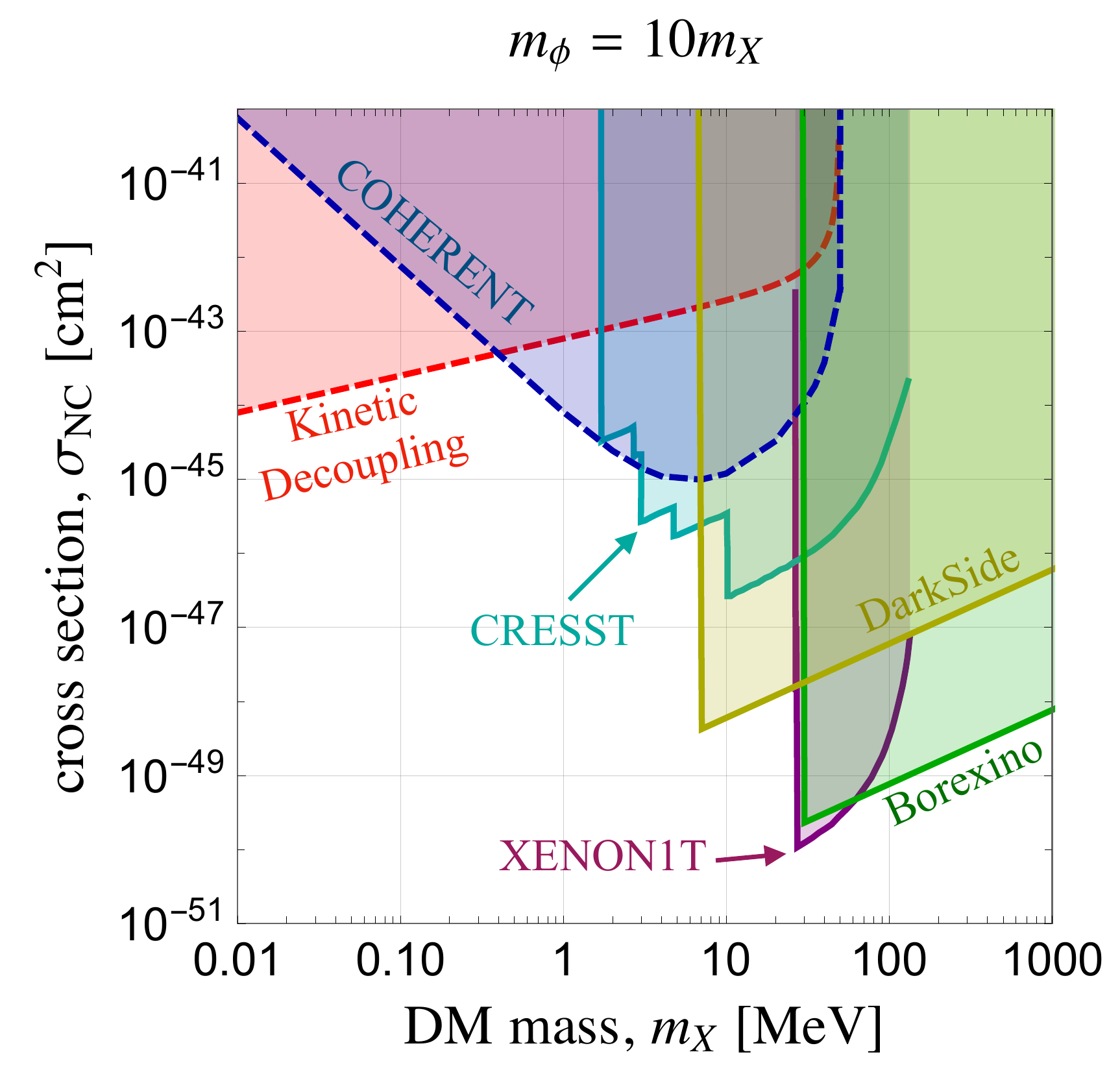}
\includegraphics[angle=0,width=.48\textwidth]{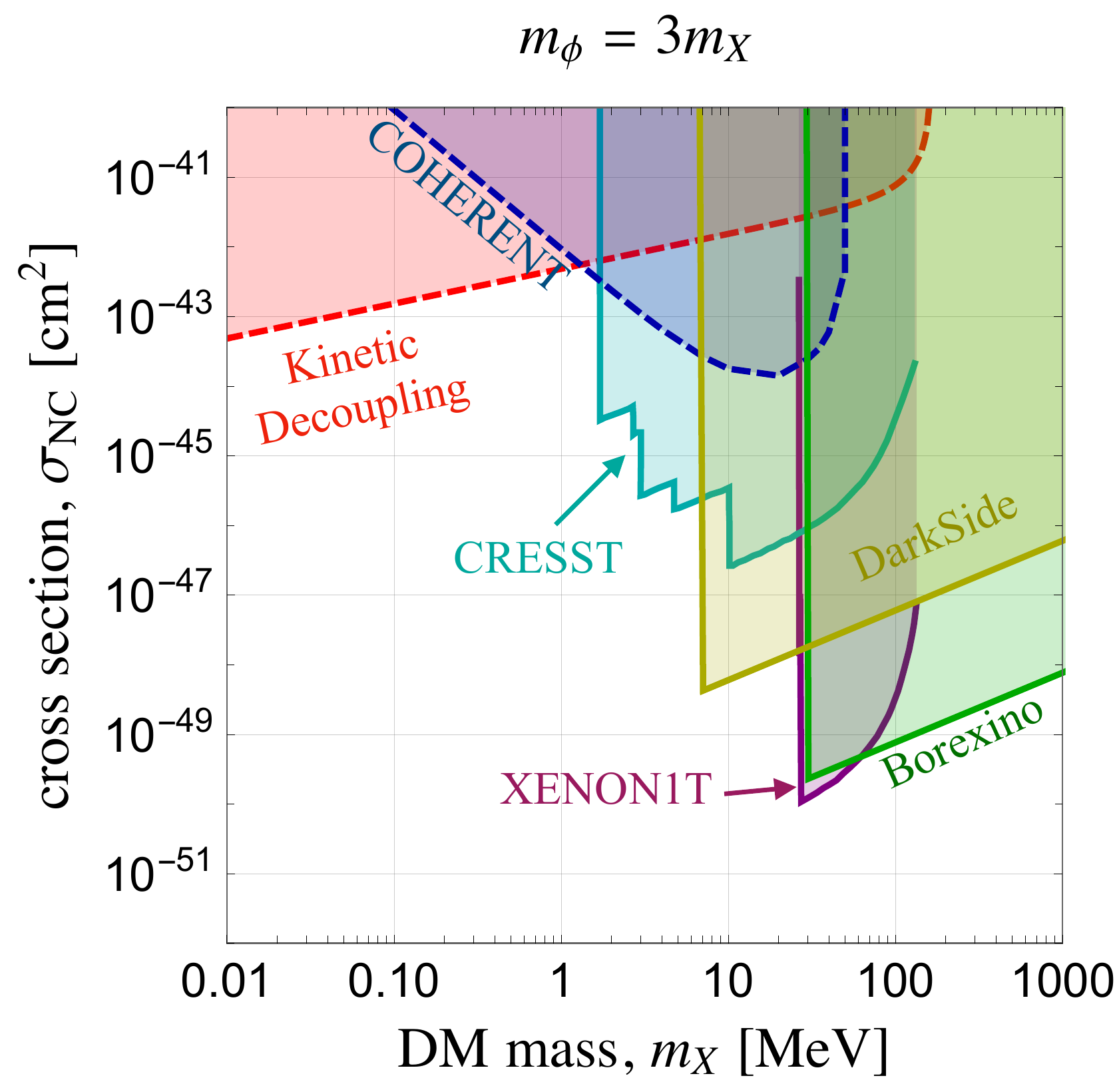}
\caption{Comparison of constraints in the $\sigma_{NC}-m_{X}$ plane from Borexino, COHERENT, CRESST, DarkSide, XENON1T and kinetic decoupling. In the left panel we fix the mediator mass to $m_{\phi}=10m_{X}$, while in the right panel $m_{\phi} = 3m_{X}$. For the kinetic decoupling curve, we assume that the nucleon-level coupling is at the largest value consistent with kaon decay constraints discussed in Sec.~\ref{sec:model}.} 
\label{fig:summary}
\end{figure*}

%
%

To date the E787 and E949 bounds on ${\rm BR} (K^{+} \rightarrow \pi^{+} \bar{\nu} \nu)$ limit the dark mediator coupling to quarks at the level 
\be y_{q} \lesssim 1.58\times 10^{-4}.
\ee
Given that this bound only constrains the $\phi$-quark coupling, it does not directly bound processes at COHERENT or direct detection, since these depend on the product $y_{q} y_{X}$. In Sec.~\ref{sec:KD} we study the kinetic decoupling of DM and neutrinos which is bounded by Lyman-alpha data. These bounds provide a constraint on $y_{X}$ and therefore in conjunction together with kaon bounds places bounds on the cross sections relevant for COHERENT and direct detection.

\subsection{Dark Matter Decay at High Masses}
Decay considerations are important for this model. In particular, the DM $X$ can decay via a 1-loop diagram $X \rightarrow \nu + \gamma$ with the rate
\be \Gamma(X \rightarrow \nu + \gamma) = \frac{y_{X}^{2}y_{q}^{2}\alpha_{EM}m_{X}^{5}}{192 \pi^{4} m_{\phi}^{4}},
\ee
where $\alpha_{EM}$ is the electromagnetic fine structure constant. 

Even the baseline requirement that DM be stable on the age of the Universe timescale leads to strong bounds on the couplings
\be
y_{X}y_{q} \lesssim 2\times 10^{-15}~\left(\frac{r}{3}\right)^{2}~\sqrt{\frac{10~{\rm MeV}}{m_{X}}},
\ee
where $r\equiv m_{\phi}/m_{X}$.

However, the minimal model sketched in Eq.~\ref{eq:model2} need not be the only source of new physics. In particular, additional new physics at higher scales can cancel the low-energy contribution in the DM decay coming from the loop of quarks. Similar arguments have been made in Ref.~\cite{Dror:2019onn}.

The decay phenomenology is similar to what is obtained in sterile neutrino dark matter models. Roughly speaking, the following mapping can be used to recast sterile neutrino DM bounds on the mixing angle: $\frac{y_{q}^{2}y_{X}^{2}}{m_{\phi}^{4}G_{F}^{2}} \lesssim \left( \sin^{2} 2 \theta \right)_{{\rm X-ray}}$.

In Ref.~\cite{Dror:2019onn} the authors comment on a $Z'$ extension with similar phenomenology. They point out that the $X \rightarrow \nu + \gamma$ is forbidden by gauge invariance. We stress that although the bounds can be strong on minimal models of this type, the bounds are alleviated at low DM mass due to the decay rate's strong DM mass dependence.

%
%


\section{Bounds from COHERENT }
\label{sec:COH}
As depicted in Fig.~\ref{fig:diagrams}(a), incoming neutrinos can be converted to DM upon scattering on a nuclear target. In this section we will describe our estimate of the sensitivity to DM from this process at the COHERENT experiment using their CsI data~\cite{Akimov:2017ade}. Similar setups have been studied previously~\cite{Farzan:2018gtr,Brdar:2018qqj}.

To lowest order, we find that the differential cross section for $\nu 
\rightarrow X$ scattering on a nucleus $N$ is
\be 
\frac{d\sigma_{\nu \rightarrow X}}{dE_{R}} =\frac{1}{4\pi}\frac{m_{N}~y_{X}^{2}y_{N}^{2}}{\left(2m_{N}E_{R} + m_{S}\right)^{2}}~\alpha
\ee
where 
\be \alpha \equiv \left(1 + \frac{E_{R}}{2m_{N}}\right)\left(\frac{m_{N}E_{R}}{E_{\nu}^{2}}+\frac{m_{X}^{2}}{2E_{\nu}^{2}}\right),
\ee
and we have introduced a nucleus-level coupling to the scalar $y_{N}$. This nucleus-level coupling can be written in terms of nucelon couplings as~\cite{Farzan:2018gtr} 
\be
y_{N} = A y_{p} + (A-Z)y_{n} 
\ee
where $y_{p},y_{n}$ are respectively the proton and neutron couplings. These nucleon couplings can finally be written in terms of quark-level couplings connecting to the Lagrangian in Eq.~\ref{eq:model2} via
\bea
y_{p} &=& m_{p} \sum_{q} \frac{y_{q}f_{q}^{p}}{m_{q}} \\
y_{n} &=& m_{n} \sum_{q} \frac{y_{q}f_{q}^{n}}{m_{q}}
\eea
where the updated scalar coefficients are taken from Ref.~\cite{Hoferichter:2015dsa}, and the quark masses from Ref.~\cite{Tanabashi:2018oca}. {Note that for equal couplings among quark flavors, $y_{p} \simeq y_{n} \simeq 17.5 y_{q}$.} In order to minimize the nuclear target dependence we follow Refs.~\cite{Farzan:2018gtr,Brdar:2018qqj} by introducing the coupling
\be
\bar{y} \equiv \sqrt{\frac{y_{N}y_{\nu}}{A}} = \sqrt{\left(\frac{Z}{A}y_{p} +\frac{A-Z}{A}y_{n}\right) }.
\ee

\begin{figure}[b!]
\includegraphics[angle=0,width=.45\textwidth]{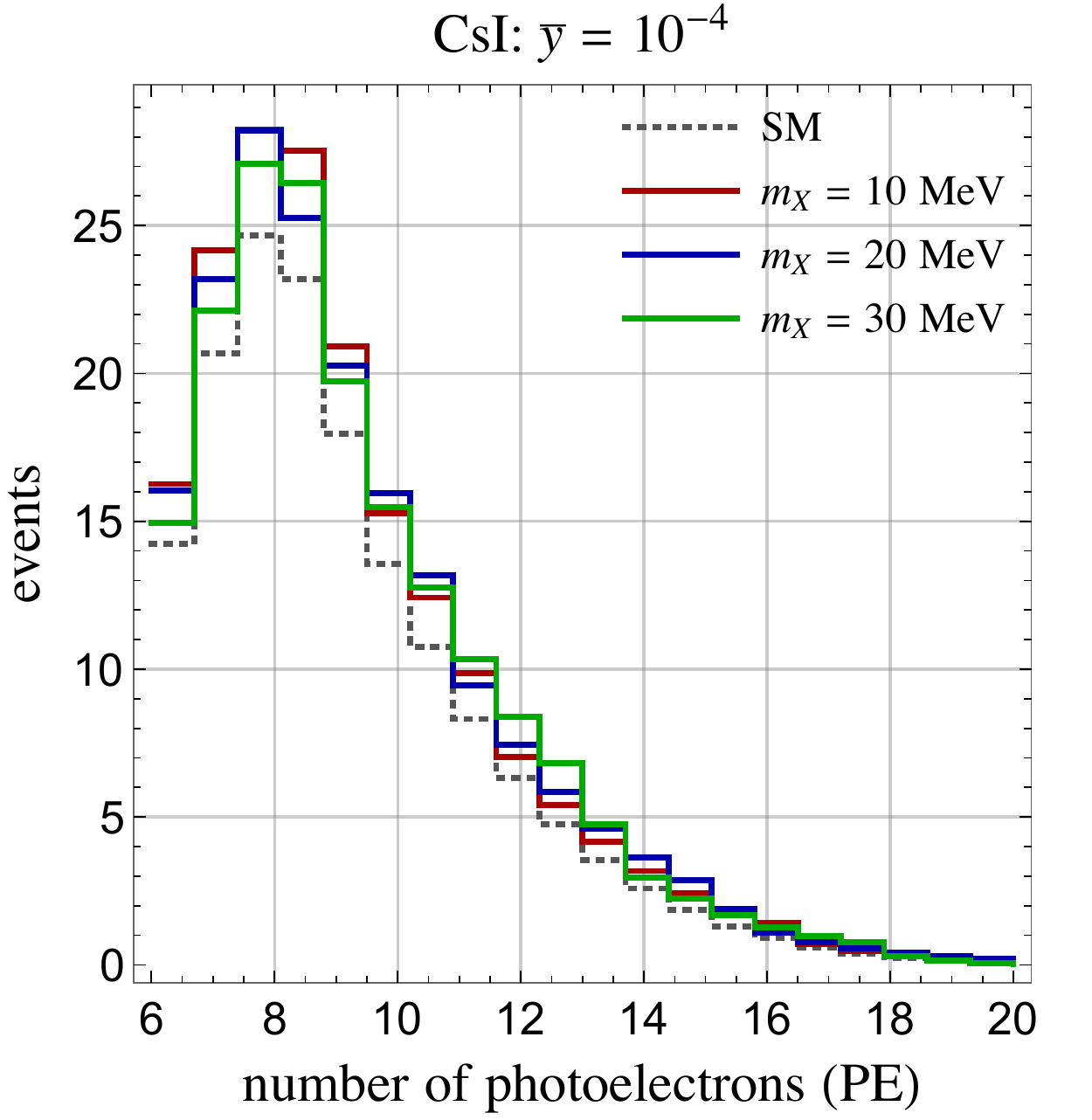}
\caption{An example spectrum comparing SM CE$\nu$NS and the $\nu\rightarrow X$ scattering spectrum for different DM masses.}
\label{fig:spectrum}
\end{figure}

The SM CE$\nu$NS rate acts as a background to this new physics search. We compute the SM event rate from the differential cross section
\be
\frac{d\sigma}{dT} = \frac{G_F^2}{4\pi}Q_w^2M \left(1-\frac{MT}{2E_\nu^2}\right)F(T)^2
\ee
where $G_{F}$ is the Fermi constant, $Q_{w} = N-(1-4\sin^{2} \theta_{w})$ is the weak nucleear hypercharge for a nucleus with $N$ neutrons and $Z$ protons, $M$ is the nuclear mass, and $F(T)$ is the nuclear form factor as a function of the recoil energy. We follow the COHERENT collaboration analysis in ~\cite{Akimov:2017ade} and use the form factor from Ref.~\cite{Klein:1999qj}. Lastly, in order to compute event rates we include the fluxes and signal acceptance function described in Ref.~\cite{Akimov:2017ade}. An example of the spectral differences that this neutrinophilic DM can induce is shown in Fig.~\ref{fig:spectrum}.

\section{Kinetic Decoupling}
\label{sec:KD}
The fact that neutrinos and the dark matter particle, $X$, share new interactions in this model has implications for small-scale structure. In particular, as long as elastic momentum-changing scattering $X + \nu \leftrightarrow X+\nu$ occur, the DM remains in approximate thermal equilibrium. Eventually these processes fail to keep up with the Hubble rate and Kinetic decoupling of DM occurs. The seeds of the first gravitationally bound DM clumps cannot form until this process ends. Thus these momentum-changing interactions effectively damp the growth of gravitational structure. After decoupling, the DM can ``freely-stream'' away from overdense potential wells. This process also erases structure on small-scales. If kinetic decoupling occurs sufficiently late it will dominate over free-streaming effects, and set the scale of the cut-off in the power spectrum~\cite{Green:2005fa,Loeb:2005pm,Bertschinger:2006nq}. Related models have been studied for their impact on late kinetic decoupling in~\cite{Bringmann:2016ilk,Binder:2016pnr}.

The damping scale in the power spectrum set by kinetic deocupling is given by the DM inside a Hubble volume,
\bea 
\label{eq:kd}
M_{{\rm cut}} &=& \frac{4\pi}{3} \frac{\rho_{DM}(T_{KD})}{H(T_{KD})} \\
&\simeq& 1.8 \times 10^{8}~M_{\odot}\left(\frac{{\rm keV}}{T_{KD}}\right)^{3},  \nonumber
\eea
where $T_{KD}$ is the temperature of kinetic decoupling.

This can be compared with constraints from Lyman-alpha data which are often quoted in terms of a constraint on the mass of warm DM. Recent data from the Lyman-alpha power spectrum require $m_{WDM}\gtrsim 5.3~{\rm keV}$~\cite{Irsic:2017ixq}. To translate to the language of kinetic decoupling, we use the fact that the cut-off induced by free-streaming is related to the WDM mass as~\cite{SommerLarsen:1999jx}
\be 
M_{f} \simeq 5.1\times 10^{10}~M_{\odot}\left(\frac{{\rm keV}}{m_{WDM}}\right)^{4}
\ee
Thus the $m_{WDM}\gtrsim 5.3~{\rm keV}$~\cite{Irsic:2017ixq} requirements implies $M_{f} < 6.5 \times10^{7}~M_{\odot}$. Using Eq~\ref{eq:kd} we see that kinetic decoupling is bounded by $T_{KD} > 1.4$ keV.


As long as $m_{X}\gtrsim 0.5$ keV, the $s$-channel resonance in $\nu$-$X$ scattering is negligible compared to the $t$-channel contribution. As is well known, at very small kinetic decoupling temperatures the damping scale set by acoustic oscillations can dominate over free-streaming effects.

Finally, we use the approximate solution of the Boltzmann equation in Ref.~\cite{Gondolo:2012vh} to solve for the temperature of kinetic decoupling by equating the Hubble rate to the momentum transfer rate $\gamma(T)$
\bea
\gamma(T) &=& \frac{1}{3m_{X}T}\int_{0}^{\infty}\frac{d^{3}p}{(2\pi)^{3}}f(p/T)(1-f(p/T))  \nonumber  \\ 
&\times & \int_{-t}^{0}dt~(-4p^{2})\frac{d\sigma_{\nu X}}{dt} 
\eea
where $t$ is the Mandelstam variable for momentum transfer and $f(p/T)$ is the Fermi-Dirac distribution. In the regime where the neutrino energy is small compared to the DM and mediator masses, this integral can be performed analytically~\cite{Bertoni:2014mva}. In this limit the kinetic decoupling temperature is found to be
\be
T_{KD} \simeq 0.4 ~{\rm keV}~\left(\frac{m_{X}}{{\rm MeV}}\right)^{5/4}\left(\frac{0.1}{y_{X}}\right)\sqrt{r^{2}-1},
\ee
where $r \equiv m_{\phi}/m_{X}$.

\begin{figure}[t!]
\includegraphics[angle=0,width=.45\textwidth]{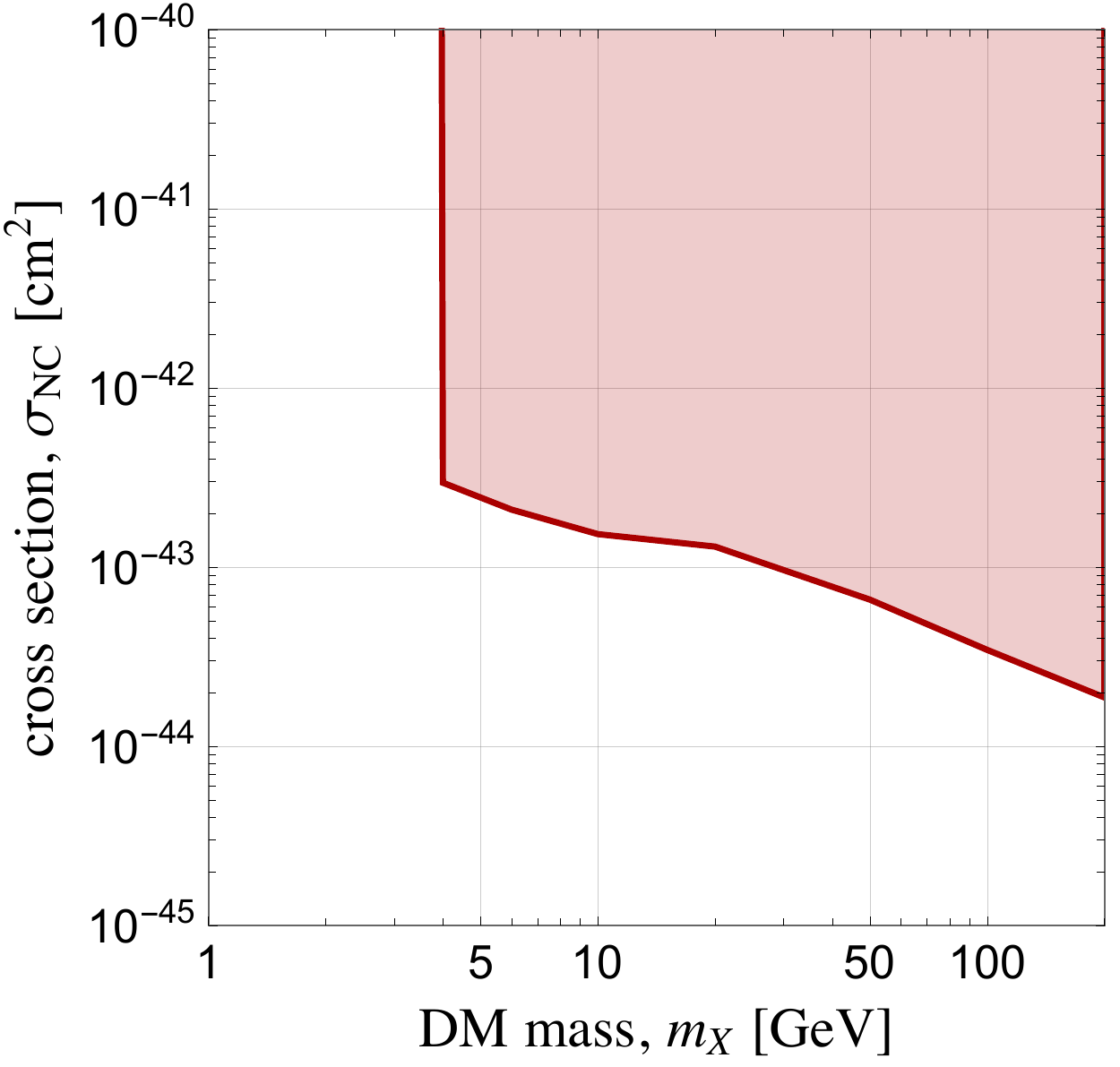}
\caption{Constraints from Super-Kamiokande~\cite{Choi:2015ara}, on the production of high-energy neutrinos from DM scattering in the Earth.}
\label{fig:thermal}
\end{figure}
\section{Bounds from Solar and Terrestrial Dark Matter Scattering}
\label{sec:SK}
Dark Matter may scatter off of a nucleus in the sun or in the Earth and convert into a neutrino as shown in Fig.~\ref{fig:summary}(b). The inelastic scattering will produce a neutrino with an energy sharply peaked around the mass of the incoming DM particle. A similar constraint was derived from the terrestrial passage of $Q$-ball dark matter in~\cite{Kusenko:2009iz}. By searching for a flux of neutrinos with this energy spectrum, constraints on the interaction cross-section can be made. As a rough estimate of the cross-section constraint, we first compute the number of neutrinos produced from the flux of DM passing through the Sun/Earth:
\be
\frac{dN_{\nu}}{dt}= \sigma_{\nu\chi} n v_{0} \rho_{\chi} V/m_{\chi};
\ee
where $n$ is the average nucleon density and $V$ is the volume of the target (either the sun or the earth) to calculate the production rate of neutrinos from this inelastic scattering process. The flux of neutrinos at Earth is then found as:
\be
F_\nu=\frac{1}{4 \pi r^2} \frac{dN_{\nu}}{dt}
\ee
where $r$ is the distance from the target at which the flux is measured.

We find that the distance of terrestrial detectors from the sun significantly reduces the flux enough that the flux from collisions in the earth is several orders of magnitude larger. Comparing the estimated neutrino flux of this novel phenomenon to bounds provided by the SuperKamiokande experiment on the tau neutrino flux~\cite{Choi:2015ara}, a cross-section on the DM inelastic scattering can be constrained in the range $4~{\rm GeV} < m_{\chi} < 200~{\rm GeV}$. Additionally, at low DM masses, the estimated flux of these neutrinos can be compared to those of solar neutrinos. The flux from this novel phenomenon for $m_{\chi} \sim 10~{\rm MeV}$, $\sigma_{\chi\nu} \sim 10^{-43}~{\rm cm^2}$ is $\sim6$ orders of magnitude lower than that of the entire Boron-8 flux of solar neutrinos, and is therefore not constrained for very low masses.


\section{Direct Detection Bounds}
\label{sec:DD}

In this model, DM direct detection recoils are highly inelastic since DM ``down-scatters'' to an essentially massless neutrino. The phenomenology of these models have been studied in~\cite{Dror:2019onn,Dror:2019dib}. As a result, the recoil spectrum is approximately given by a Dirac delta function peaked at the energy:
\be 
E_{R} = \frac{m_{X}^{2}}{2m_{N}}
\ee
where $m_{N}$ is the nuclear mass. 

With XENON1T~\cite{Aprile:2018dbl} being sensitive to $4~{\rm keV} \le E_{R} \le 40~{\rm keV}$ we therefore expect sensitivity to exist only in the following window of DM masses:
\be
30~{\rm MeV} \lesssim m_{X} \lesssim100~{\rm MeV}
\ee

Borexino can be used in a similar way to probe DM-to-neutrino down-scattering. Taking an approximate electron equivalent low-energy threshold of $\simeq 70$ keV~\cite{Bellini:2009jr}, is equivalent to a nuclear recoil threshold $E_{R} \simeq 800$ keV using the relative light output for pseudocumene~\cite{Tretyak:2009sr}. Thus for Borexino we find that scattering on the hydrogen within pseudocumene can probe DM masses down to $m_{X} \gtrsim 40~{\rm MeV}$.



We estimate compute the rate of DM-to-neutrino scattering at XENON1T and Borexino as~\cite{Dror:2019onn} 
\be 
R = \frac{\rho_{X}}{m_{X}}\sigma_{NC} \sum_{j}N_{T,j}A_{j}^{2} F_{j}^{2}~\theta(E_{R,j}^{0}-E_{{\rm th}}),
\ee
where $F$ is the nuclear form factor $N_{T}$ is the number of targets for isotope $j$, and $\sigma_{NC}$ is the neutral current cross section for DM-to-neutrino conversion.






\begin{figure}[t!]
\includegraphics[angle=0,width=.45\textwidth]{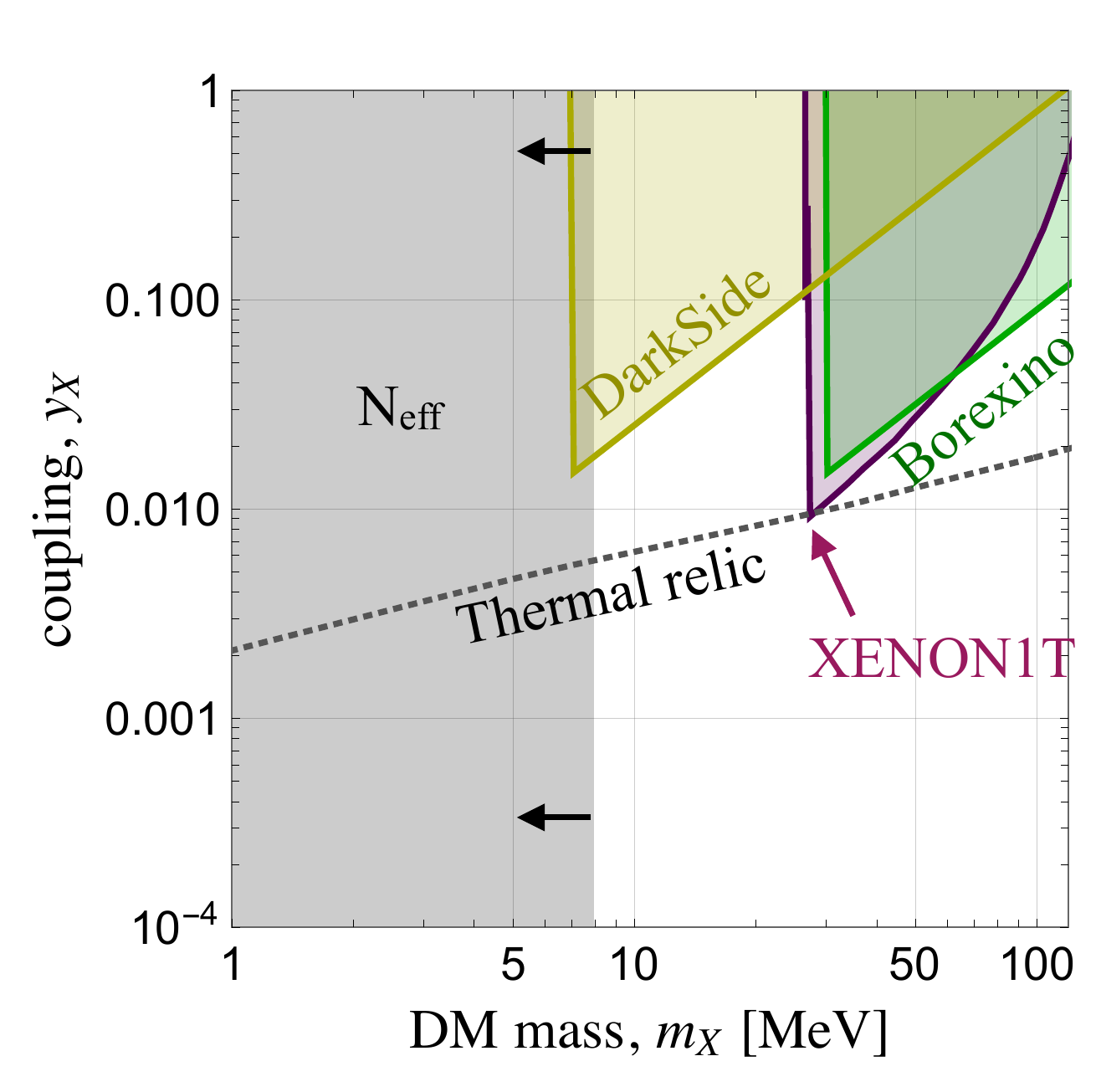}
\caption{Comparison of experiments and thermal relic prediction in coupling space. Here we have fixed $y_{q} = 10^{-9}$, consistent with kaon decay bounds. In this plot, the constraints from COHERENT on the coupling $y_{X}$ are weaker than perturbativity and thus not shown.}
\label{fig:thermal}
\end{figure}

\section{Discussion}
\label{sec:disc}

We summarize the constraints we have so far discussed in Fig.~\ref{fig:summary}, where we report bounds as a constraint on the quantity $\sigma_{NC} \equiv \frac{y_{x}^{2}y_{p}^{2}m_{X}^{2}}{16 \pi m_{\phi}^{4}}$. While the XENON1T, Borexino, DarkSide, and COHERENT bounds directly constrain the product of couplings $y_{p}y_{X}$, the kinetic decoupling constraint only places a constraint on $y_{X}$. 

We see that although direct detection and Borexino bounds dominate at high DM masses, both COHERENT and kinetic decoupling take over at low masses. Future ultra-low threshold experiments~\cite{Dror:2019onn,Dror:2019dib} may eventually be strong enough to overtake COHERENT and kinetic decoupling bounds. Finally, the constraints in Fig.~\ref{fig:summary} assume that this $\chi$ particle comprises 100\% of the dark sector mass. If $\chi$ is only a fraction of the mass, the constraints from Borexino and XENON1T will be weakened while  the COHERENT constraint would be unaffected.

Finally, we note that the thermal relic hypothesis for the dark matter abundance is realizable in this model, albeit only if the quark-level coupling is sufficiently small, $y_{q} \lesssim 10^{-9}$. In Fig.~\ref{fig:thermal} we plot the predicted $y_{X}$ coupling assuming the DM to neutrino annihilation cross section sets the relic abundance. In terms of couplings and masses, this cross section is $(\sigma v)_{\bar{X}X \rightarrow \bar{\nu} \nu} = y_{X}^{4}/(16 \pi m_{\phi}^{4})$. The other phenomenological bounds constrain the product $y_{q} y_{x}$, and thus in order to compare with the thermal relic prediction we must fix the quark-$\phi$ coupling. For illustrative purposes, we choose $y_{q} = 10^{-9}$, which yields a thermal relic solution for the DM abundance as long as DM is heavier than 8 MeV. Thermal DM lighter than this would contribute to the radiation energy density and be ruled out by $N_{{\rm eff}}$ constraints~\cite{Boehm:2013jpa}. 

\section{Conclusions}
\label{sec:conc}
The DM model described by Eq.~\ref{eq:model2} allows for a range of novel phenomenology, with interesting connections to neutrino physics. In this paper we have examined direct detection and coherent elastic neutrino nucleus scattering, finding that these two probes provide complementary constraints. At low DM masses, COHERENT provides the strongest bounds, while at larger DM masses, direct detection and Borexino prevail. We have also shown that, in this model, DM may convert into neutrinos via scattering off of nuclei in the earth, though the bounds are less competitive. Lastly, at the lowest DM masses we have found that the strongest bounds derive from late kinetic decoupling which can erase small scale structure.

Current data requires that the quark-scalar coupling be much smaller than the DM-scalar coupling. Despite these constraints on the model, the observed DM abundance can still be explained by the thermal relic hypothesis, albeit only for small quark-$\phi$ couplings.

Note that at $y_{q}= 10^{-8}$, the resultant bounds on $y_{X}$ are essentially weaker than the requirements of perturbativity. Moreover, when $y_{q} = 10^{-6}$ (or larger), there is no available parameter space consistent with the thermal relic since the COHERENT constraint rules out the 8-18 MeV mass window.

In the future, improvements in these bounds will come from ultra-low threshold direct detection~\cite{Dror:2019onn,Dror:2019dib} and CE$\nu$NS data from COHERENT, CONNIE~\cite{Aguilar-Arevalo:2016khx}, CONUS~\cite{Buck:2020opf} and $\nu$-cleus~\cite{Strauss:2017cuu}. 



{\bf \emph{Acknowledgements-  }} We are grateful to Jeff Dror and Robert McGehee for helpful conversations. This work is supported by the U.S. Department of Energy under the award number DE-SC0020250.  


\newpage

\section*{Appendix}
\section*{Freeze-out of Dark Matter in the Early Universe}
{Thermal production of DM is a well-known mechanism for obtaining the correct abundance of DM from the early Universe. In this setup, $\bar{X} X \leftrightarrow \bar{f} f$ processes keep DM in thermal equilibrium with some bath particle species $f$. Once the temperature drops below the DM mass, the abundance of DM quickly becomes Boltzmann suppressed, $\sim e^{-m_{X}/T}$. Eventually the $\bar{X} X \rightarrow \bar{f} f$ annihilation processes cease being sufficiently fast to keep up with the Hubble rate, and the DM abundance (in a comoving volume) ceases to change. This ``freeze-out'' process~\cite{Scherrer:1985zt} is one of the most studied mechanisms for the production of DM.

In our case, the annihilation process is via the annihilation to neutrinos depicted in Fig.~\ref{fig:diagrams}. Dark matter starts off in thermal equilibrium with the rest of the standard model bath which lets both the forward and backward interaction take place in the early universe. As the universe expands however, it falls out of thermal equilibrium because of lowering temperature and only the forward interaction is able to take place leading to the annihilation of dark matter particles into neutrinos. Eventually, the dark matter particles reach an equilibrium abundance which is the current dark matter abundance. 

To obtain the final DM abundance, we solved the Boltzmann equation numerically,
\be
\dot{n}_{X}+3Hn_{X}=-\langle \sigma_{ann} v \rangle (n_{X}^{2}-n_{X,EQ}^{2})
\label{eq:Bolz}
\ee
to obtain the requisite coupling $y_{X}$ needed as a function of DM mass and mediator mass. Note that in Eq.~\ref{eq:Bolz}, $\langle \sigma_{ann} v \rangle$ is the thermally averaged annihilation cross section, $H$ is the temperature-dependent Hubble rate, $n_{X}$ is the DM number density, and $n_{X,eq}$ is the equilibrium number density.   

\bibliography{crdm}

\begin{thebibliography}{50}%
\makeatletter
\providecommand \@ifxundefined [1]{%
 \@ifx{#1\undefined}
}%
\providecommand \@ifnum [1]{%
 \ifnum #1\expandafter \@firstoftwo
 \else \expandafter \@secondoftwo
 \fi
}%
\providecommand \@ifx [1]{%
 \ifx #1\expandafter \@firstoftwo
 \else \expandafter \@secondoftwo
 \fi
}%
\providecommand \natexlab [1]{#1}%
\providecommand \enquote  [1]{``#1''}%
\providecommand \bibnamefont  [1]{#1}%
\providecommand \bibfnamefont [1]{#1}%
\providecommand \citenamefont [1]{#1}%
\providecommand \href@noop [0]{\@secondoftwo}%
\providecommand \href [0]{\begingroup \@sanitize@url \@href}%
\providecommand \@href[1]{\@@startlink{#1}\@@href}%
\providecommand \@@href[1]{\endgroup#1\@@endlink}%
\providecommand \@sanitize@url [0]{\catcode `\\12\catcode `\$12\catcode
  `\&12\catcode `\#12\catcode `\^12\catcode `\_12\catcode `\%12\relax}%
\providecommand \@@startlink[1]{}%
\providecommand \@@endlink[0]{}%
\providecommand \url  [0]{\begingroup\@sanitize@url \@url }%
\providecommand \@url [1]{\endgroup\@href {#1}{\urlprefix }}%
\providecommand \urlprefix  [0]{URL }%
\providecommand \Eprint [0]{\href }%
\providecommand \doibase [0]{http://dx.doi.org/}%
\providecommand \selectlanguage [0]{\@gobble}%
\providecommand \bibinfo  [0]{\@secondoftwo}%
\providecommand \bibfield  [0]{\@secondoftwo}%
\providecommand \translation [1]{[#1]}%
\providecommand \BibitemOpen [0]{}%
\providecommand \bibitemStop [0]{}%
\providecommand \bibitemNoStop [0]{.\EOS\space}%
\providecommand \EOS [0]{\spacefactor3000\relax}%
\providecommand \BibitemShut  [1]{\csname bibitem#1\endcsname}%
\let\auto@bib@innerbib\@empty
\bibitem [{\citenamefont {Boehm}\ and\ \citenamefont
  {Fayet}(2004)}]{Boehm:2003hm}%
  \BibitemOpen
  \bibfield  {author} {\bibinfo {author} {\bibfnamefont {C.}~\bibnamefont
  {Boehm}}\ and\ \bibinfo {author} {\bibfnamefont {P.}~\bibnamefont {Fayet}},\
  }\href {\doibase 10.1016/j.nuclphysb.2004.01.015} {\bibfield  {journal}
  {\bibinfo  {journal} {Nucl. Phys. B}\ }\textbf {\bibinfo {volume} {683}},\
  \bibinfo {pages} {219} (\bibinfo {year} {2004})},\ \Eprint
  {http://arxiv.org/abs/hep-ph/0305261} {arXiv:hep-ph/0305261} \BibitemShut
  {NoStop}%
\bibitem [{\citenamefont {Hooper}\ \emph {et~al.}(2007)\citenamefont {Hooper},
  \citenamefont {Kaplinghat}, \citenamefont {Strigari},\ and\ \citenamefont
  {Zurek}}]{Hooper:2007tu}%
  \BibitemOpen
  \bibfield  {author} {\bibinfo {author} {\bibfnamefont {D.}~\bibnamefont
  {Hooper}}, \bibinfo {author} {\bibfnamefont {M.}~\bibnamefont {Kaplinghat}},
  \bibinfo {author} {\bibfnamefont {L.~E.}\ \bibnamefont {Strigari}}, \ and\
  \bibinfo {author} {\bibfnamefont {K.~M.}\ \bibnamefont {Zurek}},\ }\href
  {\doibase 10.1103/PhysRevD.76.103515} {\bibfield  {journal} {\bibinfo
  {journal} {Phys. Rev. D}\ }\textbf {\bibinfo {volume} {76}},\ \bibinfo
  {pages} {103515} (\bibinfo {year} {2007})},\ \Eprint
  {http://arxiv.org/abs/0704.2558} {arXiv:0704.2558 [astro-ph]} \BibitemShut
  {NoStop}%
\bibitem [{\citenamefont {van~den Aarssen}\ \emph {et~al.}(2012)\citenamefont
  {van~den Aarssen}, \citenamefont {Bringmann},\ and\ \citenamefont
  {Pfrommer}}]{Aarssen:2012fx}%
  \BibitemOpen
  \bibfield  {author} {\bibinfo {author} {\bibfnamefont {L.~G.}\ \bibnamefont
  {van~den Aarssen}}, \bibinfo {author} {\bibfnamefont {T.}~\bibnamefont
  {Bringmann}}, \ and\ \bibinfo {author} {\bibfnamefont {C.}~\bibnamefont
  {Pfrommer}},\ }\href {\doibase 10.1103/PhysRevLett.109.231301} {\bibfield
  {journal} {\bibinfo  {journal} {Phys. Rev. Lett.}\ }\textbf {\bibinfo
  {volume} {109}},\ \bibinfo {pages} {231301} (\bibinfo {year} {2012})},\
  \Eprint {http://arxiv.org/abs/1205.5809} {arXiv:1205.5809 [astro-ph.CO]}
  \BibitemShut {NoStop}%
\bibitem [{\citenamefont {Dasgupta}\ and\ \citenamefont
  {Kopp}(2014)}]{Dasgupta:2013zpn}%
  \BibitemOpen
  \bibfield  {author} {\bibinfo {author} {\bibfnamefont {B.}~\bibnamefont
  {Dasgupta}}\ and\ \bibinfo {author} {\bibfnamefont {J.}~\bibnamefont
  {Kopp}},\ }\href {\doibase 10.1103/PhysRevLett.112.031803} {\bibfield
  {journal} {\bibinfo  {journal} {Phys. Rev. Lett.}\ }\textbf {\bibinfo
  {volume} {112}},\ \bibinfo {pages} {031803} (\bibinfo {year} {2014})},\
  \Eprint {http://arxiv.org/abs/1310.6337} {arXiv:1310.6337 [hep-ph]}
  \BibitemShut {NoStop}%
\bibitem [{\citenamefont {Shoemaker}(2013)}]{Shoemaker:2013tda}%
  \BibitemOpen
  \bibfield  {author} {\bibinfo {author} {\bibfnamefont {I.~M.}\ \bibnamefont
  {Shoemaker}},\ }\href {\doibase 10.1016/j.dark.2013.07.002} {\bibfield
  {journal} {\bibinfo  {journal} {Phys. Dark Univ.}\ }\textbf {\bibinfo
  {volume} {2}},\ \bibinfo {pages} {157} (\bibinfo {year} {2013})},\ \Eprint
  {http://arxiv.org/abs/1305.1936} {arXiv:1305.1936 [hep-ph]} \BibitemShut
  {NoStop}%
\bibitem [{\citenamefont {Cherry}\ \emph {et~al.}(2014)\citenamefont {Cherry},
  \citenamefont {Friedland},\ and\ \citenamefont {Shoemaker}}]{Cherry:2014xra}%
  \BibitemOpen
  \bibfield  {author} {\bibinfo {author} {\bibfnamefont {J.~F.}\ \bibnamefont
  {Cherry}}, \bibinfo {author} {\bibfnamefont {A.}~\bibnamefont {Friedland}}, \
  and\ \bibinfo {author} {\bibfnamefont {I.~M.}\ \bibnamefont {Shoemaker}},\
  }\href@noop {} {\  (\bibinfo {year} {2014})},\ \Eprint
  {http://arxiv.org/abs/1411.1071} {arXiv:1411.1071 [hep-ph]} \BibitemShut
  {NoStop}%
\bibitem [{\citenamefont {Bertoni}\ \emph {et~al.}(2015)\citenamefont
  {Bertoni}, \citenamefont {Ipek}, \citenamefont {McKeen},\ and\ \citenamefont
  {Nelson}}]{Bertoni:2014mva}%
  \BibitemOpen
  \bibfield  {author} {\bibinfo {author} {\bibfnamefont {B.}~\bibnamefont
  {Bertoni}}, \bibinfo {author} {\bibfnamefont {S.}~\bibnamefont {Ipek}},
  \bibinfo {author} {\bibfnamefont {D.}~\bibnamefont {McKeen}}, \ and\ \bibinfo
  {author} {\bibfnamefont {A.~E.}\ \bibnamefont {Nelson}},\ }\href {\doibase
  10.1007/JHEP04(2015)170} {\bibfield  {journal} {\bibinfo  {journal} {JHEP}\
  }\textbf {\bibinfo {volume} {04}},\ \bibinfo {pages} {170} (\bibinfo {year}
  {2015})},\ \Eprint {http://arxiv.org/abs/1412.3113} {arXiv:1412.3113
  [hep-ph]} \BibitemShut {NoStop}%
\bibitem [{\citenamefont {Binder}\ \emph {et~al.}(2016)\citenamefont {Binder},
  \citenamefont {Covi}, \citenamefont {Kamada}, \citenamefont {Murayama},
  \citenamefont {Takahashi},\ and\ \citenamefont {Yoshida}}]{Binder:2016pnr}%
  \BibitemOpen
  \bibfield  {author} {\bibinfo {author} {\bibfnamefont {T.}~\bibnamefont
  {Binder}}, \bibinfo {author} {\bibfnamefont {L.}~\bibnamefont {Covi}},
  \bibinfo {author} {\bibfnamefont {A.}~\bibnamefont {Kamada}}, \bibinfo
  {author} {\bibfnamefont {H.}~\bibnamefont {Murayama}}, \bibinfo {author}
  {\bibfnamefont {T.}~\bibnamefont {Takahashi}}, \ and\ \bibinfo {author}
  {\bibfnamefont {N.}~\bibnamefont {Yoshida}},\ }\href {\doibase
  10.1088/1475-7516/2016/11/043} {\bibfield  {journal} {\bibinfo  {journal}
  {JCAP}\ }\textbf {\bibinfo {volume} {11}},\ \bibinfo {pages} {043} (\bibinfo
  {year} {2016})},\ \Eprint {http://arxiv.org/abs/1602.07624} {arXiv:1602.07624
  [hep-ph]} \BibitemShut {NoStop}%
\bibitem [{\citenamefont {Olivares-Del~Campo}\ \emph
  {et~al.}(2018)\citenamefont {Olivares-Del~Campo}, \citenamefont {B\oe~hm},
  \citenamefont {Palomares-Ruiz},\ and\ \citenamefont
  {Pascoli}}]{Campo:2017nwh}%
  \BibitemOpen
  \bibfield  {author} {\bibinfo {author} {\bibfnamefont {A.}~\bibnamefont
  {Olivares-Del~Campo}}, \bibinfo {author} {\bibfnamefont {C.}~\bibnamefont
  {B\oe~hm}}, \bibinfo {author} {\bibfnamefont {S.}~\bibnamefont
  {Palomares-Ruiz}}, \ and\ \bibinfo {author} {\bibfnamefont {S.}~\bibnamefont
  {Pascoli}},\ }\href {\doibase 10.1103/PhysRevD.97.075039} {\bibfield
  {journal} {\bibinfo  {journal} {Phys. Rev. D}\ }\textbf {\bibinfo {volume}
  {97}},\ \bibinfo {pages} {075039} (\bibinfo {year} {2018})},\ \Eprint
  {http://arxiv.org/abs/1711.05283} {arXiv:1711.05283 [hep-ph]} \BibitemShut
  {NoStop}%
\bibitem [{\citenamefont {Davis}\ and\ \citenamefont
  {Silk}(2015)}]{Davis:2015rza}%
  \BibitemOpen
  \bibfield  {author} {\bibinfo {author} {\bibfnamefont {J.~H.}\ \bibnamefont
  {Davis}}\ and\ \bibinfo {author} {\bibfnamefont {J.}~\bibnamefont {Silk}},\
  }\href@noop {} {\  (\bibinfo {year} {2015})},\ \Eprint
  {http://arxiv.org/abs/1505.01843} {arXiv:1505.01843 [hep-ph]} \BibitemShut
  {NoStop}%
\bibitem [{\citenamefont {Cherry}\ \emph {et~al.}(2016)\citenamefont {Cherry},
  \citenamefont {Friedland},\ and\ \citenamefont {Shoemaker}}]{Cherry:2016jol}%
  \BibitemOpen
  \bibfield  {author} {\bibinfo {author} {\bibfnamefont {J.~F.}\ \bibnamefont
  {Cherry}}, \bibinfo {author} {\bibfnamefont {A.}~\bibnamefont {Friedland}}, \
  and\ \bibinfo {author} {\bibfnamefont {I.~M.}\ \bibnamefont {Shoemaker}},\
  }\href@noop {} {\  (\bibinfo {year} {2016})},\ \Eprint
  {http://arxiv.org/abs/1605.06506} {arXiv:1605.06506 [hep-ph]} \BibitemShut
  {NoStop}%
\bibitem [{\citenamefont {Arg{\"u}elles}\ \emph {et~al.}(2017)\citenamefont
  {Arg{\"u}elles}, \citenamefont {Kheirandish},\ and\ \citenamefont
  {Vincent}}]{Arguelles:2017atb}%
  \BibitemOpen
  \bibfield  {author} {\bibinfo {author} {\bibfnamefont {C.~A.}\ \bibnamefont
  {Arg{\"u}elles}}, \bibinfo {author} {\bibfnamefont {A.}~\bibnamefont
  {Kheirandish}}, \ and\ \bibinfo {author} {\bibfnamefont {A.~C.}\ \bibnamefont
  {Vincent}},\ }\href {\doibase 10.1103/PhysRevLett.119.201801} {\bibfield
  {journal} {\bibinfo  {journal} {Phys. Rev. Lett.}\ }\textbf {\bibinfo
  {volume} {119}},\ \bibinfo {pages} {201801} (\bibinfo {year} {2017})},\
  \Eprint {http://arxiv.org/abs/1703.00451} {arXiv:1703.00451 [hep-ph]}
  \BibitemShut {NoStop}%
\bibitem [{\citenamefont {Kelly}\ and\ \citenamefont
  {Machado}(2018)}]{Kelly:2018tyg}%
  \BibitemOpen
  \bibfield  {author} {\bibinfo {author} {\bibfnamefont {K.~J.}\ \bibnamefont
  {Kelly}}\ and\ \bibinfo {author} {\bibfnamefont {P.~A.}\ \bibnamefont
  {Machado}},\ }\href {\doibase 10.1088/1475-7516/2018/10/048} {\bibfield
  {journal} {\bibinfo  {journal} {JCAP}\ }\textbf {\bibinfo {volume} {10}},\
  \bibinfo {pages} {048} (\bibinfo {year} {2018})},\ \Eprint
  {http://arxiv.org/abs/1808.02889} {arXiv:1808.02889 [hep-ph]} \BibitemShut
  {NoStop}%
\bibitem [{\citenamefont {Farzan}\ and\ \citenamefont
  {Palomares-Ruiz}(2019)}]{Farzan:2018pnk}%
  \BibitemOpen
  \bibfield  {author} {\bibinfo {author} {\bibfnamefont {Y.}~\bibnamefont
  {Farzan}}\ and\ \bibinfo {author} {\bibfnamefont {S.}~\bibnamefont
  {Palomares-Ruiz}},\ }\href {\doibase 10.1103/PhysRevD.99.051702} {\bibfield
  {journal} {\bibinfo  {journal} {Phys. Rev. D}\ }\textbf {\bibinfo {volume}
  {99}},\ \bibinfo {pages} {051702} (\bibinfo {year} {2019})},\ \Eprint
  {http://arxiv.org/abs/1810.00892} {arXiv:1810.00892 [hep-ph]} \BibitemShut
  {NoStop}%
\bibitem [{\citenamefont {Pandey}\ \emph {et~al.}(2019)\citenamefont {Pandey},
  \citenamefont {Karmakar},\ and\ \citenamefont {Rakshit}}]{Pandey:2018wvh}%
  \BibitemOpen
  \bibfield  {author} {\bibinfo {author} {\bibfnamefont {S.}~\bibnamefont
  {Pandey}}, \bibinfo {author} {\bibfnamefont {S.}~\bibnamefont {Karmakar}}, \
  and\ \bibinfo {author} {\bibfnamefont {S.}~\bibnamefont {Rakshit}},\ }\href
  {\doibase 10.1007/JHEP01(2019)095} {\bibfield  {journal} {\bibinfo  {journal}
  {JHEP}\ }\textbf {\bibinfo {volume} {01}},\ \bibinfo {pages} {095} (\bibinfo
  {year} {2019})},\ \Eprint {http://arxiv.org/abs/1810.04203} {arXiv:1810.04203
  [hep-ph]} \BibitemShut {NoStop}%
\bibitem [{\citenamefont {Choi}\ \emph {et~al.}(2019)\citenamefont {Choi},
  \citenamefont {Kim},\ and\ \citenamefont {Rott}}]{Choi:2019ixb}%
  \BibitemOpen
  \bibfield  {author} {\bibinfo {author} {\bibfnamefont {K.-Y.}\ \bibnamefont
  {Choi}}, \bibinfo {author} {\bibfnamefont {J.}~\bibnamefont {Kim}}, \ and\
  \bibinfo {author} {\bibfnamefont {C.}~\bibnamefont {Rott}},\ }\href {\doibase
  10.1103/PhysRevD.99.083018} {\bibfield  {journal} {\bibinfo  {journal} {Phys.
  Rev. D}\ }\textbf {\bibinfo {volume} {99}},\ \bibinfo {pages} {083018}
  (\bibinfo {year} {2019})},\ \Eprint {http://arxiv.org/abs/1903.03302}
  {arXiv:1903.03302 [astro-ph.CO]} \BibitemShut {NoStop}%
\bibitem [{\citenamefont {Capozzi}\ \emph {et~al.}(2017)\citenamefont
  {Capozzi}, \citenamefont {Shoemaker},\ and\ \citenamefont
  {Vecchi}}]{Capozzi:2017auw}%
  \BibitemOpen
  \bibfield  {author} {\bibinfo {author} {\bibfnamefont {F.}~\bibnamefont
  {Capozzi}}, \bibinfo {author} {\bibfnamefont {I.~M.}\ \bibnamefont
  {Shoemaker}}, \ and\ \bibinfo {author} {\bibfnamefont {L.}~\bibnamefont
  {Vecchi}},\ }\href {\doibase 10.1088/1475-7516/2017/07/021} {\bibfield
  {journal} {\bibinfo  {journal} {JCAP}\ }\textbf {\bibinfo {volume} {07}},\
  \bibinfo {pages} {021} (\bibinfo {year} {2017})},\ \Eprint
  {http://arxiv.org/abs/1702.08464} {arXiv:1702.08464 [hep-ph]} \BibitemShut
  {NoStop}%
\bibitem [{\citenamefont {Capozzi}\ \emph {et~al.}(2018)\citenamefont
  {Capozzi}, \citenamefont {Shoemaker},\ and\ \citenamefont
  {Vecchi}}]{Capozzi:2018bps}%
  \BibitemOpen
  \bibfield  {author} {\bibinfo {author} {\bibfnamefont {F.}~\bibnamefont
  {Capozzi}}, \bibinfo {author} {\bibfnamefont {I.~M.}\ \bibnamefont
  {Shoemaker}}, \ and\ \bibinfo {author} {\bibfnamefont {L.}~\bibnamefont
  {Vecchi}},\ }\href {\doibase 10.1088/1475-7516/2018/07/004} {\bibfield
  {journal} {\bibinfo  {journal} {JCAP}\ }\textbf {\bibinfo {volume} {07}},\
  \bibinfo {pages} {004} (\bibinfo {year} {2018})},\ \Eprint
  {http://arxiv.org/abs/1804.05117} {arXiv:1804.05117 [hep-ph]} \BibitemShut
  {NoStop}%
\bibitem [{\citenamefont {Scherrer}\ and\ \citenamefont
  {Turner}(1986)}]{Scherrer:1985zt}%
  \BibitemOpen
  \bibfield  {author} {\bibinfo {author} {\bibfnamefont {R.~J.}\ \bibnamefont
  {Scherrer}}\ and\ \bibinfo {author} {\bibfnamefont {M.~S.}\ \bibnamefont
  {Turner}},\ }\href {\doibase 10.1103/PhysRevD.33.1585,
  10.1103/PhysRevD.34.3263} {\bibfield  {journal} {\bibinfo  {journal} {Phys.
  Rev.}\ }\textbf {\bibinfo {volume} {D33}},\ \bibinfo {pages} {1585} (\bibinfo
  {year} {1986})},\ \bibinfo {note} {[Erratum: Phys.
  Rev.D34,3263(1986)]}\BibitemShut {NoStop}%
\bibitem [{\citenamefont {Arg{\"u}elles}\ \emph {et~al.}(2019)\citenamefont
  {Arg{\"u}elles}, \citenamefont {Diaz}, \citenamefont {Kheirandish},
  \citenamefont {Olivares-Del-Campo}, \citenamefont {Safa},\ and\ \citenamefont
  {Vincent}}]{Arguelles:2019ouk}%
  \BibitemOpen
  \bibfield  {author} {\bibinfo {author} {\bibfnamefont {C.~A.}\ \bibnamefont
  {Arg{\"u}elles}}, \bibinfo {author} {\bibfnamefont {A.}~\bibnamefont {Diaz}},
  \bibinfo {author} {\bibfnamefont {A.}~\bibnamefont {Kheirandish}}, \bibinfo
  {author} {\bibfnamefont {A.}~\bibnamefont {Olivares-Del-Campo}}, \bibinfo
  {author} {\bibfnamefont {I.}~\bibnamefont {Safa}}, \ and\ \bibinfo {author}
  {\bibfnamefont {A.~C.}\ \bibnamefont {Vincent}},\ }\href@noop {} {\
  (\bibinfo {year} {2019})},\ \Eprint {http://arxiv.org/abs/1912.09486}
  {arXiv:1912.09486 [hep-ph]} \BibitemShut {NoStop}%
\bibitem [{\citenamefont {Dror}\ \emph {et~al.}(2019)\citenamefont {Dror},
  \citenamefont {Elor},\ and\ \citenamefont {Mcgehee}}]{Dror:2019onn}%
  \BibitemOpen
  \bibfield  {author} {\bibinfo {author} {\bibfnamefont {J.~A.}\ \bibnamefont
  {Dror}}, \bibinfo {author} {\bibfnamefont {G.}~\bibnamefont {Elor}}, \ and\
  \bibinfo {author} {\bibfnamefont {R.}~\bibnamefont {Mcgehee}},\ }\href@noop
  {} {\  (\bibinfo {year} {2019})},\ \Eprint {http://arxiv.org/abs/1905.12635}
  {arXiv:1905.12635 [hep-ph]} \BibitemShut {NoStop}%
\bibitem [{\citenamefont {Dror}\ \emph {et~al.}(2020)\citenamefont {Dror},
  \citenamefont {Elor},\ and\ \citenamefont {Mcgehee}}]{Dror:2019dib}%
  \BibitemOpen
  \bibfield  {author} {\bibinfo {author} {\bibfnamefont {J.~A.}\ \bibnamefont
  {Dror}}, \bibinfo {author} {\bibfnamefont {G.}~\bibnamefont {Elor}}, \ and\
  \bibinfo {author} {\bibfnamefont {R.}~\bibnamefont {Mcgehee}},\ }\href
  {\doibase 10.1007/JHEP02(2020)134} {\bibfield  {journal} {\bibinfo  {journal}
  {JHEP}\ }\textbf {\bibinfo {volume} {02}},\ \bibinfo {pages} {134} (\bibinfo
  {year} {2020})},\ \Eprint {http://arxiv.org/abs/1908.10861} {arXiv:1908.10861
  [hep-ph]} \BibitemShut {NoStop}%
\bibitem [{\citenamefont {Brdar}\ \emph {et~al.}(2018)\citenamefont {Brdar},
  \citenamefont {Rodejohann},\ and\ \citenamefont {Xu}}]{Brdar:2018qqj}%
  \BibitemOpen
  \bibfield  {author} {\bibinfo {author} {\bibfnamefont {V.}~\bibnamefont
  {Brdar}}, \bibinfo {author} {\bibfnamefont {W.}~\bibnamefont {Rodejohann}}, \
  and\ \bibinfo {author} {\bibfnamefont {X.-J.}\ \bibnamefont {Xu}},\ }\href
  {\doibase 10.1007/JHEP12(2018)024} {\bibfield  {journal} {\bibinfo  {journal}
  {JHEP}\ }\textbf {\bibinfo {volume} {12}},\ \bibinfo {pages} {024} (\bibinfo
  {year} {2018})},\ \Eprint {http://arxiv.org/abs/1810.03626} {arXiv:1810.03626
  [hep-ph]} \BibitemShut {NoStop}%
\bibitem [{\citenamefont {Akimov}\ \emph {et~al.}(2017)\citenamefont {Akimov}
  \emph {et~al.}}]{Akimov:2017ade}%
  \BibitemOpen
  \bibfield  {author} {\bibinfo {author} {\bibfnamefont {D.}~\bibnamefont
  {Akimov}} \emph {et~al.} (\bibinfo {collaboration} {COHERENT}),\ }\href
  {\doibase 10.1126/science.aao0990} {\bibfield  {journal} {\bibinfo  {journal}
  {Science}\ }\textbf {\bibinfo {volume} {357}},\ \bibinfo {pages} {1123}
  (\bibinfo {year} {2017})},\ \Eprint {http://arxiv.org/abs/1708.01294}
  {arXiv:1708.01294 [nucl-ex]} \BibitemShut {NoStop}%
\bibitem [{\citenamefont {Aprile}\ \emph {et~al.}(2018)\citenamefont {Aprile}
  \emph {et~al.}}]{Aprile:2018dbl}%
  \BibitemOpen
  \bibfield  {author} {\bibinfo {author} {\bibfnamefont {E.}~\bibnamefont
  {Aprile}} \emph {et~al.} (\bibinfo {collaboration} {XENON}),\ }\href
  {\doibase 10.1103/PhysRevLett.121.111302} {\bibfield  {journal} {\bibinfo
  {journal} {Phys. Rev. Lett.}\ }\textbf {\bibinfo {volume} {121}},\ \bibinfo
  {pages} {111302} (\bibinfo {year} {2018})},\ \Eprint
  {http://arxiv.org/abs/1805.12562} {arXiv:1805.12562 [astro-ph.CO]}
  \BibitemShut {NoStop}%
\bibitem [{\citenamefont {Bellini}\ \emph {et~al.}(2013)\citenamefont {Bellini}
  \emph {et~al.}}]{Bellini:2013uui}%
  \BibitemOpen
  \bibfield  {author} {\bibinfo {author} {\bibfnamefont {G.}~\bibnamefont
  {Bellini}} \emph {et~al.} (\bibinfo {collaboration} {Borexino}),\ }\href
  {\doibase 10.1103/PhysRevD.88.072010} {\bibfield  {journal} {\bibinfo
  {journal} {Phys. Rev.}\ }\textbf {\bibinfo {volume} {D88}},\ \bibinfo {pages}
  {072010} (\bibinfo {year} {2013})},\ \Eprint {http://arxiv.org/abs/1311.5347}
  {arXiv:1311.5347 [hep-ex]} \BibitemShut {NoStop}%
\bibitem [{\citenamefont {Krnjaic}(2016)}]{Krnjaic:2015mbs}%
  \BibitemOpen
  \bibfield  {author} {\bibinfo {author} {\bibfnamefont {G.}~\bibnamefont
  {Krnjaic}},\ }\href {\doibase 10.1103/PhysRevD.94.073009} {\bibfield
  {journal} {\bibinfo  {journal} {Phys. Rev.}\ }\textbf {\bibinfo {volume}
  {D94}},\ \bibinfo {pages} {073009} (\bibinfo {year} {2016})},\ \Eprint
  {http://arxiv.org/abs/1512.04119} {arXiv:1512.04119 [hep-ph]} \BibitemShut
  {NoStop}%
\bibitem [{\citenamefont {Batell}\ \emph {et~al.}(2019)\citenamefont {Batell},
  \citenamefont {Freitas}, \citenamefont {Ismail},\ and\ \citenamefont
  {Mckeen}}]{Batell:2018fqo}%
  \BibitemOpen
  \bibfield  {author} {\bibinfo {author} {\bibfnamefont {B.}~\bibnamefont
  {Batell}}, \bibinfo {author} {\bibfnamefont {A.}~\bibnamefont {Freitas}},
  \bibinfo {author} {\bibfnamefont {A.}~\bibnamefont {Ismail}}, \ and\ \bibinfo
  {author} {\bibfnamefont {D.}~\bibnamefont {Mckeen}},\ }\href {\doibase
  10.1103/PhysRevD.100.095020} {\bibfield  {journal} {\bibinfo  {journal}
  {Phys. Rev. D}\ }\textbf {\bibinfo {volume} {100}},\ \bibinfo {pages}
  {095020} (\bibinfo {year} {2019})},\ \Eprint
  {http://arxiv.org/abs/1812.05103} {arXiv:1812.05103 [hep-ph]} \BibitemShut
  {NoStop}%
\bibitem [{\citenamefont {Artamonov}\ \emph {et~al.}(2008)\citenamefont
  {Artamonov} \emph {et~al.}}]{Artamonov:2008qb}%
  \BibitemOpen
  \bibfield  {author} {\bibinfo {author} {\bibfnamefont {A.~V.}\ \bibnamefont
  {Artamonov}} \emph {et~al.} (\bibinfo {collaboration} {E949}),\ }\href
  {\doibase 10.1103/PhysRevLett.101.191802} {\bibfield  {journal} {\bibinfo
  {journal} {Phys. Rev. Lett.}\ }\textbf {\bibinfo {volume} {101}},\ \bibinfo
  {pages} {191802} (\bibinfo {year} {2008})},\ \Eprint
  {http://arxiv.org/abs/0808.2459} {arXiv:0808.2459 [hep-ex]} \BibitemShut
  {NoStop}%
\bibitem [{\citenamefont {Cortina~Gil}\ \emph {et~al.}(2019)\citenamefont
  {Cortina~Gil} \emph {et~al.}}]{CortinaGil:2018fkc}%
  \BibitemOpen
  \bibfield  {author} {\bibinfo {author} {\bibfnamefont {E.}~\bibnamefont
  {Cortina~Gil}} \emph {et~al.} (\bibinfo {collaboration} {NA62}),\ }\href
  {\doibase 10.1016/j.physletb.2019.01.067} {\bibfield  {journal} {\bibinfo
  {journal} {Phys. Lett.}\ }\textbf {\bibinfo {volume} {B791}},\ \bibinfo
  {pages} {156} (\bibinfo {year} {2019})},\ \Eprint
  {http://arxiv.org/abs/1811.08508} {arXiv:1811.08508 [hep-ex]} \BibitemShut
  {NoStop}%
\bibitem [{\citenamefont {Lurkin}(2019)}]{Lurkin:2019brq}%
  \BibitemOpen
  \bibfield  {author} {\bibinfo {author} {\bibfnamefont {N.}~\bibnamefont
  {Lurkin}} (\bibinfo {collaboration} {NA62}),\ }in\ \href@noop {} {\emph
  {\bibinfo {booktitle} {{An Alpine LHC Physics Summit 2019 (ALPS 2019)
  Obergurgl, Austria, April 22-27, 2019}}}}\ (\bibinfo {year} {2019})\ \Eprint
  {http://arxiv.org/abs/1907.12955} {arXiv:1907.12955 [hep-ex]} \BibitemShut
  {NoStop}%
\bibitem [{\citenamefont {Farzan}\ \emph {et~al.}(2018)\citenamefont {Farzan},
  \citenamefont {Lindner}, \citenamefont {Rodejohann},\ and\ \citenamefont
  {Xu}}]{Farzan:2018gtr}%
  \BibitemOpen
  \bibfield  {author} {\bibinfo {author} {\bibfnamefont {Y.}~\bibnamefont
  {Farzan}}, \bibinfo {author} {\bibfnamefont {M.}~\bibnamefont {Lindner}},
  \bibinfo {author} {\bibfnamefont {W.}~\bibnamefont {Rodejohann}}, \ and\
  \bibinfo {author} {\bibfnamefont {X.-J.}\ \bibnamefont {Xu}},\ }\href
  {\doibase 10.1007/JHEP05(2018)066} {\bibfield  {journal} {\bibinfo  {journal}
  {JHEP}\ }\textbf {\bibinfo {volume} {05}},\ \bibinfo {pages} {066} (\bibinfo
  {year} {2018})},\ \Eprint {http://arxiv.org/abs/1802.05171} {arXiv:1802.05171
  [hep-ph]} \BibitemShut {NoStop}%
\bibitem [{\citenamefont {Hoferichter}\ \emph {et~al.}(2015)\citenamefont
  {Hoferichter}, \citenamefont {Ruiz~de Elvira}, \citenamefont {Kubis},\ and\
  \citenamefont {Mei{\ss}ner}}]{Hoferichter:2015dsa}%
  \BibitemOpen
  \bibfield  {author} {\bibinfo {author} {\bibfnamefont {M.}~\bibnamefont
  {Hoferichter}}, \bibinfo {author} {\bibfnamefont {J.}~\bibnamefont {Ruiz~de
  Elvira}}, \bibinfo {author} {\bibfnamefont {B.}~\bibnamefont {Kubis}}, \ and\
  \bibinfo {author} {\bibfnamefont {U.-G.}\ \bibnamefont {Mei{\ss}ner}},\
  }\href {\doibase 10.1103/PhysRevLett.115.092301} {\bibfield  {journal}
  {\bibinfo  {journal} {Phys. Rev. Lett.}\ }\textbf {\bibinfo {volume} {115}},\
  \bibinfo {pages} {092301} (\bibinfo {year} {2015})},\ \Eprint
  {http://arxiv.org/abs/1506.04142} {arXiv:1506.04142 [hep-ph]} \BibitemShut
  {NoStop}%
\bibitem [{\citenamefont {Tanabashi}\ \emph {et~al.}(2018)\citenamefont
  {Tanabashi} \emph {et~al.}}]{Tanabashi:2018oca}%
  \BibitemOpen
  \bibfield  {author} {\bibinfo {author} {\bibfnamefont {M.}~\bibnamefont
  {Tanabashi}} \emph {et~al.} (\bibinfo {collaboration} {Particle Data
  Group}),\ }\href {\doibase 10.1103/PhysRevD.98.030001} {\bibfield  {journal}
  {\bibinfo  {journal} {Phys. Rev.}\ }\textbf {\bibinfo {volume} {D98}},\
  \bibinfo {pages} {030001} (\bibinfo {year} {2018})}\BibitemShut {NoStop}%
\bibitem [{\citenamefont {Klein}\ and\ \citenamefont
  {Nystrand}(1999)}]{Klein:1999qj}%
  \BibitemOpen
  \bibfield  {author} {\bibinfo {author} {\bibfnamefont {S.}~\bibnamefont
  {Klein}}\ and\ \bibinfo {author} {\bibfnamefont {J.}~\bibnamefont
  {Nystrand}},\ }\href {\doibase 10.1103/PhysRevC.60.014903} {\bibfield
  {journal} {\bibinfo  {journal} {Phys. Rev. C}\ }\textbf {\bibinfo {volume}
  {60}},\ \bibinfo {pages} {014903} (\bibinfo {year} {1999})},\ \Eprint
  {http://arxiv.org/abs/hep-ph/9902259} {arXiv:hep-ph/9902259} \BibitemShut
  {NoStop}%
\bibitem [{\citenamefont {Green}\ \emph {et~al.}(2005)\citenamefont {Green},
  \citenamefont {Hofmann},\ and\ \citenamefont {Schwarz}}]{Green:2005fa}%
  \BibitemOpen
  \bibfield  {author} {\bibinfo {author} {\bibfnamefont {A.~M.}\ \bibnamefont
  {Green}}, \bibinfo {author} {\bibfnamefont {S.}~\bibnamefont {Hofmann}}, \
  and\ \bibinfo {author} {\bibfnamefont {D.~J.}\ \bibnamefont {Schwarz}},\
  }\href {\doibase 10.1088/1475-7516/2005/08/003} {\bibfield  {journal}
  {\bibinfo  {journal} {JCAP}\ }\textbf {\bibinfo {volume} {08}},\ \bibinfo
  {pages} {003} (\bibinfo {year} {2005})},\ \Eprint
  {http://arxiv.org/abs/astro-ph/0503387} {arXiv:astro-ph/0503387} \BibitemShut
  {NoStop}%
\bibitem [{\citenamefont {Loeb}\ and\ \citenamefont
  {Zaldarriaga}(2005)}]{Loeb:2005pm}%
  \BibitemOpen
  \bibfield  {author} {\bibinfo {author} {\bibfnamefont {A.}~\bibnamefont
  {Loeb}}\ and\ \bibinfo {author} {\bibfnamefont {M.}~\bibnamefont
  {Zaldarriaga}},\ }\href {\doibase 10.1103/PhysRevD.71.103520} {\bibfield
  {journal} {\bibinfo  {journal} {Phys. Rev. D}\ }\textbf {\bibinfo {volume}
  {71}},\ \bibinfo {pages} {103520} (\bibinfo {year} {2005})},\ \Eprint
  {http://arxiv.org/abs/astro-ph/0504112} {arXiv:astro-ph/0504112} \BibitemShut
  {NoStop}%
\bibitem [{\citenamefont {Bertschinger}(2006)}]{Bertschinger:2006nq}%
  \BibitemOpen
  \bibfield  {author} {\bibinfo {author} {\bibfnamefont {E.}~\bibnamefont
  {Bertschinger}},\ }\href {\doibase 10.1103/PhysRevD.74.063509} {\bibfield
  {journal} {\bibinfo  {journal} {Phys. Rev. D}\ }\textbf {\bibinfo {volume}
  {74}},\ \bibinfo {pages} {063509} (\bibinfo {year} {2006})},\ \Eprint
  {http://arxiv.org/abs/astro-ph/0607319} {arXiv:astro-ph/0607319} \BibitemShut
  {NoStop}%
\bibitem [{\citenamefont {Bringmann}\ \emph {et~al.}(2016)\citenamefont
  {Bringmann}, \citenamefont {Ihle}, \citenamefont {Kersten},\ and\
  \citenamefont {Walia}}]{Bringmann:2016ilk}%
  \BibitemOpen
  \bibfield  {author} {\bibinfo {author} {\bibfnamefont {T.}~\bibnamefont
  {Bringmann}}, \bibinfo {author} {\bibfnamefont {H.~T.}\ \bibnamefont {Ihle}},
  \bibinfo {author} {\bibfnamefont {J.}~\bibnamefont {Kersten}}, \ and\
  \bibinfo {author} {\bibfnamefont {P.}~\bibnamefont {Walia}},\ }\href
  {\doibase 10.1103/PhysRevD.94.103529} {\bibfield  {journal} {\bibinfo
  {journal} {Phys. Rev. D}\ }\textbf {\bibinfo {volume} {94}},\ \bibinfo
  {pages} {103529} (\bibinfo {year} {2016})},\ \Eprint
  {http://arxiv.org/abs/1603.04884} {arXiv:1603.04884 [hep-ph]} \BibitemShut
  {NoStop}%
\bibitem [{\citenamefont {Ir\v~si\v c}\ \emph {et~al.}(2017)\citenamefont
  {Ir\v~si\v c} \emph {et~al.}}]{Irsic:2017ixq}%
  \BibitemOpen
  \bibfield  {author} {\bibinfo {author} {\bibfnamefont {V.}~\bibnamefont
  {Ir\v~si\v c}} \emph {et~al.},\ }\href {\doibase 10.1103/PhysRevD.96.023522}
  {\bibfield  {journal} {\bibinfo  {journal} {Phys. Rev. D}\ }\textbf {\bibinfo
  {volume} {96}},\ \bibinfo {pages} {023522} (\bibinfo {year} {2017})},\
  \Eprint {http://arxiv.org/abs/1702.01764} {arXiv:1702.01764 [astro-ph.CO]}
  \BibitemShut {NoStop}%
\bibitem [{\citenamefont {Sommer-Larsen}\ and\ \citenamefont
  {Dolgov}(2001)}]{SommerLarsen:1999jx}%
  \BibitemOpen
  \bibfield  {author} {\bibinfo {author} {\bibfnamefont {J.}~\bibnamefont
  {Sommer-Larsen}}\ and\ \bibinfo {author} {\bibfnamefont {A.}~\bibnamefont
  {Dolgov}},\ }\href {\doibase 10.1086/320211} {\bibfield  {journal} {\bibinfo
  {journal} {Astrophys. J.}\ }\textbf {\bibinfo {volume} {551}},\ \bibinfo
  {pages} {608} (\bibinfo {year} {2001})},\ \Eprint
  {http://arxiv.org/abs/astro-ph/9912166} {arXiv:astro-ph/9912166} \BibitemShut
  {NoStop}%
\bibitem [{\citenamefont {Gondolo}\ \emph {et~al.}(2012)\citenamefont
  {Gondolo}, \citenamefont {Hisano},\ and\ \citenamefont
  {Kadota}}]{Gondolo:2012vh}%
  \BibitemOpen
  \bibfield  {author} {\bibinfo {author} {\bibfnamefont {P.}~\bibnamefont
  {Gondolo}}, \bibinfo {author} {\bibfnamefont {J.}~\bibnamefont {Hisano}}, \
  and\ \bibinfo {author} {\bibfnamefont {K.}~\bibnamefont {Kadota}},\ }\href
  {\doibase 10.1103/PhysRevD.86.083523} {\bibfield  {journal} {\bibinfo
  {journal} {Phys. Rev. D}\ }\textbf {\bibinfo {volume} {86}},\ \bibinfo
  {pages} {083523} (\bibinfo {year} {2012})},\ \Eprint
  {http://arxiv.org/abs/1205.1914} {arXiv:1205.1914 [hep-ph]} \BibitemShut
  {NoStop}%
\bibitem [{\citenamefont {Choi}\ \emph {et~al.}(2015)\citenamefont {Choi} \emph
  {et~al.}}]{Choi:2015ara}%
  \BibitemOpen
  \bibfield  {author} {\bibinfo {author} {\bibfnamefont {K.}~\bibnamefont
  {Choi}} \emph {et~al.} (\bibinfo {collaboration} {Super-Kamiokande}),\ }\href
  {\doibase 10.1103/PhysRevLett.114.141301} {\bibfield  {journal} {\bibinfo
  {journal} {Phys. Rev. Lett.}\ }\textbf {\bibinfo {volume} {114}},\ \bibinfo
  {pages} {141301} (\bibinfo {year} {2015})},\ \Eprint
  {http://arxiv.org/abs/1503.04858} {arXiv:1503.04858 [hep-ex]} \BibitemShut
  {NoStop}%
\bibitem [{\citenamefont {Kusenko}\ and\ \citenamefont
  {Shoemaker}(2009)}]{Kusenko:2009iz}%
  \BibitemOpen
  \bibfield  {author} {\bibinfo {author} {\bibfnamefont {A.}~\bibnamefont
  {Kusenko}}\ and\ \bibinfo {author} {\bibfnamefont {I.~M.}\ \bibnamefont
  {Shoemaker}},\ }\href {\doibase 10.1103/PhysRevD.80.027701} {\bibfield
  {journal} {\bibinfo  {journal} {Phys. Rev. D}\ }\textbf {\bibinfo {volume}
  {80}},\ \bibinfo {pages} {027701} (\bibinfo {year} {2009})},\ \Eprint
  {http://arxiv.org/abs/0905.3929} {arXiv:0905.3929 [hep-ph]} \BibitemShut
  {NoStop}%
\bibitem [{\citenamefont {Bellini}\ \emph {et~al.}(2010)\citenamefont {Bellini}
  \emph {et~al.}}]{Bellini:2009jr}%
  \BibitemOpen
  \bibfield  {author} {\bibinfo {author} {\bibfnamefont {G.}~\bibnamefont
  {Bellini}} \emph {et~al.} (\bibinfo {collaboration} {Borexino}),\ }\href
  {\doibase 10.1103/PhysRevC.81.034317} {\bibfield  {journal} {\bibinfo
  {journal} {Phys. Rev. C}\ }\textbf {\bibinfo {volume} {81}},\ \bibinfo
  {pages} {034317} (\bibinfo {year} {2010})},\ \Eprint
  {http://arxiv.org/abs/0911.0548} {arXiv:0911.0548 [hep-ex]} \BibitemShut
  {NoStop}%
\bibitem [{\citenamefont {Tretyak}(2010)}]{Tretyak:2009sr}%
  \BibitemOpen
  \bibfield  {author} {\bibinfo {author} {\bibfnamefont {V.}~\bibnamefont
  {Tretyak}},\ }\href {\doibase 10.1016/j.astropartphys.2009.11.002} {\bibfield
   {journal} {\bibinfo  {journal} {Astropart. Phys.}\ }\textbf {\bibinfo
  {volume} {33}},\ \bibinfo {pages} {40} (\bibinfo {year} {2010})},\ \Eprint
  {http://arxiv.org/abs/0911.3041} {arXiv:0911.3041 [nucl-ex]} \BibitemShut
  {NoStop}%
\bibitem [{\citenamefont {Boehm}\ \emph {et~al.}(2013)\citenamefont {Boehm},
  \citenamefont {Dolan},\ and\ \citenamefont {McCabe}}]{Boehm:2013jpa}%
  \BibitemOpen
  \bibfield  {author} {\bibinfo {author} {\bibfnamefont {C.}~\bibnamefont
  {Boehm}}, \bibinfo {author} {\bibfnamefont {M.~J.}\ \bibnamefont {Dolan}}, \
  and\ \bibinfo {author} {\bibfnamefont {C.}~\bibnamefont {McCabe}},\ }\href
  {\doibase 10.1088/1475-7516/2013/08/041} {\bibfield  {journal} {\bibinfo
  {journal} {JCAP}\ }\textbf {\bibinfo {volume} {08}},\ \bibinfo {pages} {041}
  (\bibinfo {year} {2013})},\ \Eprint {http://arxiv.org/abs/1303.6270}
  {arXiv:1303.6270 [hep-ph]} \BibitemShut {NoStop}%
\bibitem [{\citenamefont {Aguilar-Arevalo}\ \emph {et~al.}(2016)\citenamefont
  {Aguilar-Arevalo} \emph {et~al.}}]{Aguilar-Arevalo:2016khx}%
  \BibitemOpen
  \bibfield  {author} {\bibinfo {author} {\bibfnamefont {A.}~\bibnamefont
  {Aguilar-Arevalo}} \emph {et~al.} (\bibinfo {collaboration} {CONNIE}),\
  }\href {\doibase 10.1088/1742-6596/761/1/012057} {\bibfield  {journal}
  {\bibinfo  {journal} {J. Phys. Conf. Ser.}\ }\textbf {\bibinfo {volume}
  {761}},\ \bibinfo {pages} {012057} (\bibinfo {year} {2016})},\ \Eprint
  {http://arxiv.org/abs/1608.01565} {arXiv:1608.01565 [physics.ins-det]}
  \BibitemShut {NoStop}%
\bibitem [{\citenamefont {Buck}\ \emph {et~al.}(2020)\citenamefont {Buck} \emph
  {et~al.}}]{Buck:2020opf}%
  \BibitemOpen
  \bibfield  {author} {\bibinfo {author} {\bibfnamefont {C.}~\bibnamefont
  {Buck}} \emph {et~al.},\ }\href {\doibase 10.1088/1742-6596/1342/1/012094}
  {\bibfield  {journal} {\bibinfo  {journal} {J. Phys. Conf. Ser.}\ }\textbf
  {\bibinfo {volume} {1342}},\ \bibinfo {pages} {012094} (\bibinfo {year}
  {2020})}\BibitemShut {NoStop}%
\bibitem [{\citenamefont {Strauss}\ \emph {et~al.}(2017)\citenamefont {Strauss}
  \emph {et~al.}}]{Strauss:2017cuu}%
  \BibitemOpen
  \bibfield  {author} {\bibinfo {author} {\bibfnamefont {R.}~\bibnamefont
  {Strauss}} \emph {et~al.},\ }\href {\doibase 10.1140/epjc/s10052-017-5068-2}
  {\bibfield  {journal} {\bibinfo  {journal} {Eur. Phys. J. C}\ }\textbf
  {\bibinfo {volume} {77}},\ \bibinfo {pages} {506} (\bibinfo {year} {2017})},\
  \Eprint {http://arxiv.org/abs/1704.04320} {arXiv:1704.04320
  [physics.ins-det]} \BibitemShut {NoStop}%
\end{thebibliography}%

\end{document}